\documentstyle[12pt,epsf]{article}

\def\Epara#1#2{{\cal E}^{\ {#1}}_{\parallel\ {#2}}}
\def\Eperp#1#2{{\cal E}^{\ {#1}}_{\perp\, {#2}}}

\begin{document}

\begin{titlepage}
\begin{flushright}
  KUNS-1621\\[-1mm]
  {\tt hep-ph/9912496}
\end{flushright}

\vspace*{5ex}
\begin{center}
{\large\bf
  Probing extra dimensions using Nambu-Goldstone bosons}
\vspace{5ex}

Taichiro Kugo\footnote{E-mail: kugo@gauge.scphys.kyoto-u.ac.jp}
and 
Koichi Yoshioka\footnote{E-mail: yoshioka@gauge.scphys.kyoto-u.ac.jp}

\vspace{1ex}

{\it
Department of Physics, Kyoto University, Kyoto 606-8502, Japan
}

\end{center}
\vspace*{8ex}

\begin{abstract}
We investigate a possibility that our four-dimensional world is a
brane-like object embedded in a higher dimensional spacetime. In such
a situation, the transverse coordinates of the brane become the
Nambu-Goldstone bosons which appear as a result of spontaneous
breaking of the translation symmetry. We determine the form of the
effective action of the system, finding the explicit form of the
vierbein induced on the brane in terms of the Nambu-Goldstone boson
variables and the bulk vielbein. As was pointed out in the previous
paper, the Kaluza-Klein mode couplings are suppressed by the effect of
the brane fluctuation and the suppression is stronger if the brane
tension is smaller. However, we here show that the brane tension
cannot be arbitrarily small since the inverse of the brane tension
gives the coupling constant of the Nambu-Goldstone bosons. A rather
stringent bound is obtained for the brane tension and the 
fundamental (`string') scale from the consideration of the cooling
process of the supernova.
\end{abstract}
\end{titlepage}
\setcounter{footnote}{0}

\section{Introduction}
\setcounter{equation}{0}

To understand the existence of various hierarchical scales in nature
is one of the most important problem in particle physics. Recently it
was proposed in~\cite{hierarchy} that the existence of extra spacetime
dimensions may solve the gauge hierarchy problem between the Planck
and weak scales in four dimensions. Stimulated by their work, there
have appeared a large number of papers concerning the physics of extra
dimensions. They deal with various phenomenological problems, such as
fermion mass hierarchy~\cite{fermion}, neutrino 
physics~\cite{neutrino}, supersymmetry breaking~\cite{susy}, flavor 
problems~\cite{flavor}, cosmology and astrophysics~\cite{astro}, and
so on. 

It is really an interesting possibility that our
four-dimensional world may lie on a `brane' like a D-brane,
orientifold plane, domain wall, etc.\ embedded in larger 
dimensions~\cite{brane}. What are then typical signatures for such a
brane world? Since there cannot exist a rigid body in the relativistic
theory, any type of brane must necessarily fluctuate. Therefore there
are always scalar fields which stand for the coordinates of the brane
in the transverse dimensions. These scalar fields are the
Nambu-Goldstone (NG) bosons which appear as a result of spontaneous
breaking of the translation symmetry by the presence of the
brane. Then one important way to explore the possibility of brane
world is to investigate the effects of this NG bosons.

In higher dimensional models where the gravity and/or gauge fields
live in the bulk, there are infinite numbers of Kaluza-Klein (KK)
gravitons and/or KK gauge bosons on the brane. They couple to the
matters living on the brane (our world) and can give visible effects
in future collider experiments and astrophysics. These effects have
been intensively investigated so far in many 
articles~\cite{KKgravity}. However, in calculating some processes with
KK excitation modes, there was a puzzling problem. That is, in summing
up infinitely many KK modes, a {\it tree-level} amplitude diverges
when the number of extra dimensions is larger than or equal to 2. In
the previous work~\cite{BKNY}, we showed that the higher KK mode
couplings to the fields confined on the brane are exponentially
suppressed if the effect of the above mentioned NG bosons is properly
taken into account. That is, the suppression is due to the fluctuation
of the brane. The suppression factor works as a regulator in the
summations of KK mode contributions and cures the problematic
divergences. We also discussed some important phenomenological
consequences of this factor. There, we concluded that if the brane
tension is very small, the KK mode contributions become completely
invisible from our world.

In this paper, we will consider the effects of other interactions of
this NG field and then show that we cannot freely have a small value
for the brane tension. For this purpose, we will examine two physical
phenomena. One is a new long-range force mediated by the NG bosons and
the other is the cooling process of neutron stars in the supernova
explosions. Both results involve the brane tension parameter $f$ (the
decay constant of the NG boson) in the form $1/f^8$ and lead to
sizable effects if the tension becomes very small. That is, when the
brane tension becomes small the KK mode couplings are suppressed but
instead the effects of the NG bosons can be measured. Therefore, we do
not miss the possibilities of finding signatures of the extra
dimensions in near future.

To discuss the effects of the NG bosons, we need explicitly construct
the effective action on the brane. The construction of the action is
more or less similar to that in the string theory, and is
straightforward in principle, as was done 
in Ref.~\cite{sundrum}. Although the author of Ref.~\cite{sundrum} has
given a definition of the induced vielbein, he unfortunately gave only
a perturbative procedure for finding its form in terms of the bulk
gravity and the brane coordinates (NG fields). Since it seems to have
appeared in no literature, we give an explicit expression for the
induced vielbein in this paper. The details of the derivation of the
explicit form, together with its relevance to the non-linear
realization theory, are given in Appendix A.

This paper is organized as follows. In section 2, we give the
effective action for the system of the $d$-dimensional brane embedded
in a higher dimensional spacetime where the gravity and gauge fields
live in the bulk and the matter fields only in the brane. The NG boson
part of this action is discussed in some detail in section 3. Given
those interactions in the flat background, we consider several
phenomenological effects of the NG field in section 4. Comparing
them with observations, we can obtain constraints for physical
parameters of extra dimensions such as the brane tension stated
above. Section 5 is devoted to summary of this paper. Appendixes B and 
C are added for the calculations of the fifth force potential and of
the cross sections for the two processes relevant to the star
cooling.

\section{Setup}
\setcounter{equation}{0}

The brane action is basically well-known in string
theory~\cite{Pol}. So we here explain the setup briefly which we use
throughout this paper, and construct the action for the system, in
particular, by presenting the explicit form of the induced vierbein.

We take the brane to be 
a $(d-1)$-brane whose world volume topology is $R^d$. It is embedded
in the bulk spacetime of dimension $D (>d)$ with 
topology $R^d\times T^{D-d}$, where $T^k$ denotes a $k$-dimensional
torus. The coordinates of the bulk spacetime are denoted $X^M$, and
those of the brane world volume are denoted $x^\mu$. The indices of
upper-case Roman letters from the middle, $M,\,N,\cdots,$ run over all
dimensions, $0,\,1,\cdots,(D-1)$ and the Greek letter 
indices $\mu,\,\nu,\cdots,$ run only over the first $d$ 
dimensions, $0,\,1,\cdots,(d-1)$, while the lower-case Roman 
letters $m,\,n,\cdots$ run over the 
rest $(D-d)$-dimensions, $d,\cdots,(D-1)$; namely, we write 
like $M =(\mu,m),\,N=(\nu,n),\cdots$. The local Lorentz indices are
denoted similarly by the corresponding letters from the beginning of
the alphabet, like $A=(\alpha,a),\,B=(\beta,b),\cdots$. Our notation
is summarized in table 1.

\begin{table}[htbp]
  \begin{center}
    \begin{tabular}{l|c|c} \hline
                 & bulk spacetime & brane world volume \\ \hline
      coordinate & $X^M\;\; (M=0,1,\cdots,D-1)$ & $x^\mu\;\;
      (\mu=0,1,\cdots,d-1)$ \\ \hline 
    \end{tabular} 
    \vspace{4mm}

    \begin{tabular}{l|c||c|c} \hline
      & $0,\,1,\cdots,(D-1)$ & $0,\,1,\cdots,(d-1)$ 
      & $d,\,d+1,\cdots,(D-1)$ \\ \hline 
      curved indices & $M,\,N,\cdots$ & $\mu,\,\nu,\cdots$
      & $m,\,n,\cdots$ \\ \hline
      local Lorentz indices & $A,\,B,\cdots$ & $\alpha,\,\beta,\cdots$
      & $a,\,b,\cdots$ \\ \hline
    \end{tabular} 
    \caption{Summary of our notation.}
  \end{center}
\end{table}

The action of this system consists of two parts; the bulk part
$S_{\rm bulk}$ and the brane part $S_{\rm brane}$. The bulk part
action takes the form
\begin{eqnarray}
  S_{\rm bulk} &=& \int d^DX \,\det{E} \left[-\Lambda
    +\frac{M^{D-2}}{2} R -\frac{1}{4}G^{MR}G^{NS} {\rm tr}
    (F_{MN}F_{RS}) +\cdots\right],
  \label{Sbulk}
\end{eqnarray}
where $\Lambda, M$ and $R$ are the cosmological constant, the
$D$-dimensional fundamental scale, and the $D$-dimensional scalar
curvature, respectively. We have shown only gravity and Yang-Mills
terms explicitly, $E^A_{\ M}(X)$ is the vielbein and $A_M(X)$ the
Yang-Mills fields. The bulk metric $G_{MN}(X)$ is given by the
vielbein as usual
\begin{eqnarray}
  G_{MN}(X) &=& \eta_{AB} E^A_{\ M}(X)E^B_{\ N}(X),
\end{eqnarray}
where $\eta_{AB}$ is the $D$-dimensional Minkowski metric: $\eta_{AB}=
{\rm diag} (+1,-1,\cdots,-1)$. The inverse matrices of
$G_{MN}$ and $E^A_{\ \,M}$ are denoted $G^{MN}$ and $E^M_{\ \ A}$,
respectively. 

Now, let $Y^M(x)$ denote a point in the bulk spacetime which the 
point $x$ on the brane occupies. Considering the distance between two
points $x$ and $x+dx$ on the brane and the parallel transport of a
charged field from $x$ to $x+dx$, we see that the following metric and
gauge fields are induced on the brane from the bulk ones, $G_{MN}(X)$
and $A_M(X)$:
\begin{eqnarray}
  g_{\mu\nu}(x) &=& G_{MN}(Y(x))\partial_\mu Y^M \partial_\nu Y^N,
  \nonumber \\
  a_{\mu}(x) &=& A_{M}(Y(x))\partial_\mu Y^M.
  \label{inducedYM}
\end{eqnarray}
Similarly, a vielbein $e^\alpha_{\ \mu}(x)$ on the brane is also
induced from the bulk one $E^A_{\ M}(X)$. The definition of this
induced vielbein is, however, a bit non-trivial problem as was
discussed by Sundrum~\cite{sundrum} and hence is explained in detail
in Appendix A\@. We derive there an explicit form of the induced
vielbein, which we need to discuss the phenomenological implications
of the NG bosons. The explicit form is given by
\begin{eqnarray}
  e^\alpha_{\ \mu}(x) &=& 
  \Epara{\alpha}{\nu} \left(1 +({\cal E}^{\rm T}_{\parallel}\eta\,  
    {\cal E}_{\parallel})^{-1} ({\cal E}^{\rm T}_{\perp} \eta\, 
    {\cal E}_{\perp})\right)^{1/2\,\nu}_{\ \ \ \ \ \mu}, 
  \label{inducedVB}
\end{eqnarray}
where we have defined
\begin{eqnarray}
  \pmatrix{\Epara{\alpha}{\mu}(x) \cr \Eperp{a}{\mu}(x) \cr} &\equiv& 
  \pmatrix{E^\alpha_{\ M}(Y(x))\,\partial_\mu Y^M \cr 
    E^a_{\ M}(Y(x))\,\partial_\mu Y^M},
  \label{calY}
\end{eqnarray}
and $({\cal E}^{\rm T}_{\parallel}\eta\, 
{\cal E}_{\parallel})_{\mu\nu}= 
\Epara{\alpha}{\mu}\eta_{\alpha\beta}\Epara{\beta}{\nu}$, 
$({\cal E}^{\rm T}_{\perp}\eta\, {\cal E}_{\perp})_{\mu\nu}=
\Eperp{a}{\mu}\eta_{ab}\Eperp{b}{\nu}$ 
and $({\cal E}^{\rm T}_{\parallel}\eta\, 
{\cal E}_{\parallel})^{-1\,\mu\nu}$ is the inverse matrix of 
$({\cal E}^{\rm T}_{\parallel}\eta\, 
{\cal E}_{\parallel})_{\mu\nu}$. This vielbein satisfies the
condition, $g_{\mu\nu}(x)=\eta_{\alpha\beta}\, 
e^\alpha_\mu(x) e^\beta_\nu(x)$ and gives a matrix connecting the
local Lorentz basis to the curved one on the brane. The spin
connection $\omega_{\mu\ \,\beta}^{\ \,\alpha}(x)$ induced on the
brane from the bulk one $\Omega_{M\ \,B}^{\ \ \,A}(X)$ is also given
in Appendix A\@. It becomes the usual connection $\omega_\mu(e)$ given
by the induced vielbein $e^\alpha_{\ \mu}$ if the bulk connection is
the usual one $\Omega_M(E)$ by $E^A_{\ \,M}$.

The fields living on the brane generally couple to the bulk fields 
through these induced vielbein (or metric) and Yang-Mills fields. 
Let $\psi(x)$ be a fermion field on the brane which is charged under
the Yang-Mills gauge group. Then, the brane part action takes the form
\begin{equation}
  S_{\rm brane} \,=\, \int d^dx \det{e} \biggl[-\tau 
  +e^\mu_{\ \alpha}(x) \bar\psi(x) i\gamma^\alpha
  \Bigl(\frac{\stackrel{\leftrightarrow}{\nabla}_\mu}{2} -ig 
  a_\mu(x)\Bigr) \psi(x) -m\bar\psi(x)\psi(x) +\cdots\, \biggr],
  \label{Sbrane}
\end{equation}
where $\bar\psi{\stackrel{\leftrightarrow}{\nabla}_\mu} \psi\equiv 
\bar\psi\nabla_\mu\psi-(\nabla_\mu\bar\psi) \psi$ 
with $\nabla_\mu\psi=(\partial_\mu+{i\over 4}
\omega_\mu^{\ \alpha\beta}(x)\sigma_{\alpha\beta})\psi$. The first
term is a cosmological constant term on the brane with $\tau$ standing
for the brane tension, and it also gives the Nambu-Goto action
determining the motion (fluctuation) of the brane. The ellipsis
contains other bosonic part terms for scalars and 
gauge fields if there exist such fields living only on the
brane. These terms can be easily written by using the induced 
metric $g_{\mu\nu}(x)$. It should be noted that the total system
with the action $S=S_{\rm bulk}+S_{\rm brane}$ still keeps the gauge
invariance under the bulk general coordinate, local Lorentz and
Yang-Mills gauge transformations. Under the bulk general coordinate
transformation, the induced fields $e^\mu_{\ \alpha}(x), a_\mu(x)$ as
well as the genuine field $\psi(x)$ on the brane, all transform as
scalar fields. This can be checked easily from the 
equations (\ref{inducedYM})--(\ref{calY}) if we note 
that $\partial_\mu Y^M(x)$ transforms as bulk vectors. The bulk local
Lorentz transformation results in an $SO(1,d-1)$ local Lorentz
transformation on the induced vielbein $e^\alpha_{\ \mu}(x)$ on the
brane (see Appendix A), but $S_{\rm brane}$ is manifestly invariant
under any $SO(1,d-1)$ local Lorentz transformation. Moreover, under
the bulk Yang-Mills transformation $A'_M(X)=U(X)A_M(X)U^{-1}(X)
+(i/g)U(X)\partial_MU^{-1}(X)$, the induced gauge field $a_\mu(x)$ is
transformed as
\begin{eqnarray}
  a'_\mu(x) &=& U(Y(x))a_\mu(x)U^{-1}(Y(x)) +(i/g)U(Y(x))\partial_\mu
  U^{-1}(Y(x)).
\end{eqnarray}
This is the usual gauge transformation in $d$-dimensions 
with $U(x)=U(Y(x))$, under which $S_{\rm brane}$ is manifestly
invariant.

The brane action $S_{\rm brane}$ also has an invariance under
reparametrization of the world volume coordinates $x^\mu$ as is
familiar in string theory~\cite{Pol}. We fix this reparametrization
invariance by choosing the gauge condition (the static gauge)
\begin{eqnarray}
  Y^{M=\mu}(x) &=& x^\mu.
  \label{reparaGF}
\end{eqnarray}
Namely, we identify the world volume coordinates $x^\mu$ with the
first $d$ components of the brane coordinate $Y^M(x)$ in the
bulk. Then, the remaining $(D-d)$ components $Y^{M=m}(x)$ represent
the brane coordinates transverse to the brane and behave as dynamical
scalar fields on the brane. They are the NG fields appearing as a
result of spontaneous breaking of the translation symmetries
transverse to the brane.

If we take a suitable gauge fixing also for the local Lorentz
invariance in the bulk, the induced vielbein can be written in a
simpler form. Let us take the following local Lorentz gauge as is
usually adopted in the case of dimensional reduction:
\begin{eqnarray}
  E_{\ M}^{A} &=& \pmatrix{
    E_{\ \mu}^\alpha & E_{\ m}^\alpha=0 \cr
    E_{\ \mu}^a\equiv E_{\ m}^a B^m_\mu & E_{\ m}^a \cr}, \\[1mm]
  E_{\ A}^{M} &=& \pmatrix{
    E_{\ \alpha}^\mu & E_{\ a}^\mu=0 \cr
    E_{\ \alpha}^m = -B^m_\mu E^\mu_{\ \alpha} & E_{\ a}^m \cr},
\end{eqnarray}
which can be realized by using local Lorentz $SO(1,D-1)/SO(1,d-1)$
transformations. Note that $E_{\ \alpha}^\mu$ and $E_{\ a}^m$ are the
inverse matrices of $E_{\ \mu}^\alpha$ and $E_{\ m}^a$,
respectively. In this local Lorentz gauge (and in the static gauge),
we have
\begin{eqnarray}
  \Epara{\alpha}{\mu}(x) &=& E_{\ \mu}^\alpha (x,Y^m(x)), \\
  \Eperp{a}{\mu}(x) &=& E_{\ n}^a(x,Y^m(x)) {\cal B}_\mu^n(x),\\[2mm]
  {\cal B}_\mu^n(x) &\equiv& B_\mu^n(x,Y^m(x)) +\partial_\mu Y^n(x),
\end{eqnarray}
so that from (\ref{inducedVB}) the induced vielbein becomes 
\begin{eqnarray}
  e^\alpha_{\ \mu}(x) &=& E_{\ \nu}^\alpha \left(\delta^\nu_{\ \mu}
    +G^{\nu\rho} {\cal B}_\rho^m G_{mn} {\cal B}_\mu^n \right)^{1/2}.
  \label{VB}
\end{eqnarray}
In the usual Kaluza-Klein gravity, the zero modes of the off-diagonal
components of the higher dimensional metric, $B^m_\mu$, become
massless gauge bosons of $U(1)^{D-d}$ in $d$ dimensions. They
correspond to the gauged translational symmetry in the extra
dimension. Now, since the translations are spontaneously broken by the
existence of the brane, these gauge fields absorb the NG bosons $Y^m$
and become massive on the brane. In the above expression of the
induced vielbein (\ref{VB}), we can see a part of this Higgs
effect. That is, the gauge fields $B^m_\mu$ appear only in the form 
of ${\cal B}^m_\mu=B^m_\mu+\partial_\mu Y^m$ which correspond to the
massive gauge fields.

It is interesting to note that even if the bulk gravity is absent,
i.e.,
\begin{eqnarray}
  E^A_{\ M}(X) \;=\; \delta^A_{\ M} && (G_{MN}(X) \;=\; \eta_{MN}),
\end{eqnarray}
the induced vielbein $e^\alpha_{\ \mu}(x)$ is 
non-trivial; $e^\alpha_{\ \mu}(x)\neq \delta^\alpha_{\ \mu}$. In this
case, the Eq.~(\ref{VB}) takes a simpler form
\begin{eqnarray}
  e^\alpha_{\ \mu}(x) &=& \left(\delta^\alpha_{\ \mu}
    -\frac{1}{\tau}\partial^\alpha\phi^m(x)\partial_\mu\phi^m(x)
  \right)^{1/2},  
  \label{vielbein}
\end{eqnarray}
where $\phi^m(x)$ are $(D-d)$ NG scalar fields rescaled with the brane
tension factor so that they carry the usual mass dimension $(d/2-1)$
of scalar fields;
\begin{eqnarray}
  \phi^m(x) &\equiv& \sqrt{\tau} Y^m(x). 
\end{eqnarray}
Hereafter, we consider the flat background case, i.e., there is no
object in the bulk except for the brane. We take into account the
fluctuations of metric around this background. In this situation, the
expression for $S_{\rm bulk}$ after the torus compactification and the 
relevant phenomenology have been discussed in detail in 
Refs.~\cite{KKgravity}. In the following part of this paper, we will
investigate the physics with the 
action $S=S_{\rm bulk}+S_{\rm brane}$, focusing especially on the
role of the NG field $\phi^m(x)$.

\section{Effective action on the brane}
\setcounter{equation}{0}

The fluctuation of brane is governed by the Nambu-Goto term in the
brane action Eq.~(\ref{Sbrane}), which reads by inserting the form of 
the induced vierbein (\ref{vielbein}),
\begin{eqnarray}
  \int d^dx \det{e}\, (-\tau) &=& \int d^dx\, \left[-\tau +\frac{1}{2}
    \partial^\mu\phi^m(x)\partial_\mu\phi^m(x)
    +\frac{1}{8\tau}(\partial^\mu\phi^m(x)\partial_\mu\phi^m(x))^2
    \nonumber \right. \\
  &&\left.\qquad -\frac{1}{4\tau}(\partial^\mu\phi^m(x)
    \partial_\nu\phi^m(x)) (\partial^\nu\phi^n(x)
    \partial_\mu\phi^n(x)) +\cdots \right].
\end{eqnarray}
Note that $\phi$ has a properly normalized kinetic term by the
rescaling $Y^m(x)=\phi^m(x)/\sqrt{\tau}$, and hence that all
interaction terms of $\phi(x)$ are accompanied by some powers 
of $1/\tau$. Noting also that the induced Yang-Mills 
field (\ref{inducedYM}) now reads
\begin{eqnarray}
  a_\mu(x) &=& A_\mu \Bigl(x,\frac{\phi(x)}{\sqrt{\tau}}\Bigr)
  +\frac{1}{\sqrt{\tau}} A_m\Bigl(x,\frac{\phi(x)}{\sqrt{\tau}}\Bigr)
  \partial_\mu\phi^m(x), 
  \label{inducedA}
\end{eqnarray}
the fermion part of the brane action is given by 
\begin{eqnarray}
  S_{\rm fermion} &=& \int d^dx\,\bar\psi(i\gamma^\mu\partial_\mu
  -m_\psi) \psi +S_{\rm NG} + S_{\rm grav},
\end{eqnarray}
\begin{eqnarray}
  S_{\rm NG} &\!=\!& \int d^dx\,\biggl[\,g\bar\psi\gamma^\mu\psi A_\mu
  \Bigl(x,\frac{\phi(x)}{\sqrt{\tau}}\Bigr) +\frac{g}{\sqrt{\tau}}
  \bar\psi\gamma^\mu\psi A_m \Bigl(x,\frac{\phi(x)}{\sqrt{\tau}}\Bigr)
  \partial_\mu\phi^m(x) \nonumber \biggr.\\
  &&\biggl.\quad +\frac{1}{2\tau}
  (\partial^\mu\phi^m\partial_\nu\phi^m -\delta^\mu_{\ \nu} 
  \partial^\rho\phi^m\partial_\rho\phi^m) \Bigl(\bar\psi i\gamma^\nu
  {\frac{\stackrel{\leftrightarrow}{\partial}_\mu}{2}}\psi 
  +g\bar\psi\gamma^\nu\psi A_\mu \Bigl(x,\frac{\phi(x)}{\sqrt{\tau}}
  \Bigr)\Bigr) \biggr. \nonumber \\[1mm]
  &&\biggl.\quad +\frac{1}{2\tau}
  (\partial^\rho\phi^m\partial_\rho\phi^m) m_\psi\bar\psi\psi
  +O(\tau^{-\frac{3}{2}})\, \biggr],
  \label{SNG} \\[2mm]
  S_{\rm grav} &=& -\kappa \int d^dx\,
  h^{\mu\nu}(x,\frac{\phi(x)}{\sqrt{\tau}})\, T_{\mu\nu},
  \label{Sgrav}
\end{eqnarray}
where $h_{\mu\nu}$ is $d$-dimensional part of the fluctuations of the
vielbein, $E^\alpha_\mu=\delta^\alpha_\mu+\kappa h^\alpha_\mu$, and
$\kappa$ denotes the $d$-dimensional gravity coupling constant which 
is related to the fundamental scale $M$ in (\ref{Sbulk}) via
$M^{D-2}V^{D-d}=\kappa^{-2}$. $V$ is the volume of the
compactification manifold $T^{D-d}$ which is $(2\pi R)^{D-d}$ here and 
when $d=4$, $\kappa$ is related to the Newton constant $G_N$ by
$\kappa^2=8\pi G_N$. The energy-momentum tensor $T_{\mu\nu}$ is
given by
\begin{eqnarray}
  T_{\mu\nu} &=& -\eta_{\mu\nu} \Bigl(\,\frac{1}{2}\bar\psi
  i\gamma^\rho {\stackrel{\leftrightarrow}{\partial}_\rho}\psi-m_\psi
  \bar\psi\psi \Bigr) +\frac{1}{2}\bar\psi i\gamma_\nu
  {\stackrel{\leftrightarrow}{\partial}_\mu}\psi +\frac{1}{2}
  \partial_\rho(\bar\psi i\gamma_\mu\sigma_\nu^{\ \,\rho}\psi),
\end{eqnarray}
assuming $h_{\mu\nu}$ symmetric. Here we have neglected the
interaction terms between the fluctuations of bulk metric and the NG
field, which are higher-order terms in the gravitational coupling
constant and $\tau^{-1}$.

From the above action, we can see that there are two types of coupling
between the NG boson $\phi$ and the fields on the brane. One is the
derivative couplings of $\phi$ (the second and third lines 
in (\ref{SNG})) which come from the expansion of the induced 
vielbein (\ref{vielbein}). The derivative coupling is governed 
by $\tau$ which is the decay constant of this NG field. Another type
of interaction originates from the fact that $\phi$ stand for the
coordinates of brane in the transverse dimensions. To see the form of
this type of coupling, we ignore all the derivative interaction terms
of $\phi$ through $\partial_\mu\phi$. Then, the brane system is simply
described by
\begin{eqnarray}
  S_{\rm brane} &=& \int d^dx\,\left[\frac{1}{2}\partial^\mu\phi^m
    \partial_\mu\phi^m +\bar\psi(i\gamma^\mu\partial_\mu-m_\psi)\psi
    +g\bar\psi\gamma^\mu\psi(x)A_\mu
    \Bigl(x,\frac{\phi(x)}{\sqrt{\tau}}\Bigr) \right],
\end{eqnarray}
aside from the gravity coupling terms. We assume, for simplicity, that 
the extra dimensions are all compactified into a tori with a common
radius $R$. Then the bulk gauge fields $A_M(X)$ is expanded into the
KK modes labeled by a $(D-d)$ vector $n=(n^m)$:
\begin{eqnarray}
  A_M(X^\mu=x^\mu,X^m=Y^m) &=& \frac{1}{\sqrt{V}} \sum_n A_M^{(n)}(x)
  e^{in\cdot Y/R}.
\end{eqnarray}
Inserting the Kaluza-Klein mode expansion into $A_\mu$, the gauge
interaction term reads 
\begin{eqnarray}
  \int d^dx \sum_{n} g\bar\psi(x)\gamma^\mu\psi(x)A^{(n)}_\mu(x)
  \exp\Bigl(\frac{in\cdot\phi(x)}{R\sqrt{\tau}}\Bigr).
\end{eqnarray}
This interaction term apparently seems to imply equal couplings $g$
for all the Kaluza-Klein excited modes. However, although $\phi=0$
classically it has fluctuations quantum mechanically, which is
governed by the kinetic 
term $(1/2)\partial^\mu\phi^m\partial_\mu\phi^m$. We should rewrite
the exponential factor into a normal ordered form
\begin{eqnarray}
  \int d^dx\sum_{n} g\, e^{-\frac{1}{2}\frac{n^2}{R^2\tau}
  \Delta(M^{-1})}\cdot \bar\psi(x)\gamma^\mu\psi(x)A^{(n)}_\mu(x)
  :\exp\Bigl(\frac{in\cdot\phi(x)}{R\sqrt{\tau}}\Bigr):\,,
  \label{normal}
\end{eqnarray}
where $\Delta$ is the free propagator of $\phi$
\begin{eqnarray}
  \Delta(x-y) &\equiv& \langle \phi(x)\phi(y) \rangle \;=\;
  \frac{-1}{4\pi^2} \frac{1}{(x-y)^2}.
\end{eqnarray}
In deriving this, a singularity at $x=y$ is cut off at the fundamental
scale $M$ above which the effective theory description on the brane
becomes invalid. The interaction term (\ref{normal}) implies that the
effective coupling of the level $n$ KK modes to $d$-dimensional fields
is suppressed exponentially. It should be noted that the same
exponential factor also appears in the coupling with the KK 
gravitons $S_{\rm grav}$ (\ref{Sgrav}). In the following section, we
consider the effects of the NG boson $\phi(x)$ with the derivative
interaction terms in (\ref{SNG}) and non-derivative 
one (\ref{normal}). We will see that these two interaction terms
play complementary roles in obtaining the limits for the parameters of
the model.

Here we comment on the coupling involving the bulk gauge filed $A_m$
in the action (\ref{SNG}). When we perform a compactification
from higher $D$-dimensional theories, the KK modes $A_\mu^{(n)}$ for
the bulk gauge fields get masses and so should absorb the physical
degrees of freedom from some scalar fields. These are just supplied by 
the KK excitations $A_m^{(n)}$ of the extra-dimensional components;
i.e., the induced gauge field (\ref{inducedA}) can be rewritten in the 
form
\begin{eqnarray}
  a_\mu(x) &=& \sum_{n\neq 0} \left[ \widetilde A_\mu^{(n)}(x)
    e^{i\frac{n}{R}\frac{\phi(x)}{\sqrt{\tau}}} -i\frac{R}{n^2}
    \partial_\mu \Bigl(n^m A_m^{(n)}(x)
    e^{i\frac{n}{R}\frac{\phi(x)}{\sqrt{\tau}}} \Bigr)\right] 
  +({\rm zero~ mode}),
\end{eqnarray}
where $\widetilde A_\mu^{(n)}\equiv A_\mu^{(n)} +i(R/n^2)\partial_\mu 
(n^m A_m^{(n)})$ are massive gauge fields on the brane (with 
mass $|n/R|$) which are invariant (covariant in the non-abelian case)
under the gauge transformation in $D$ dimensions. Since the induced
gauge fields only appear in the minimal interactions to brane fields,
the second term can be absorbed by a redefinition of those fields (a
gauge transformation). In the following, the Kaluza-Klein gauge
fields $A_\mu^{(n)}$ should be understood as this massive gauge 
field $\widetilde A_\mu^{(n)}$. It should be noted that in the above
equation, the zero mode part contains the zero modes of extra 
components $A_m$ which remains massless. Since such massless `scalar'
fields in the adjoint representation have not been experimentally
observed, they must be removed from the low-energy effective
theories. Of course, it is not necessary that gauge fields are living
in the bulk unlike the graviton. The effects of the NG field, which we
will discuss below, actually exist even when the gauge fields are
confined on the brane to start with. However, if one wish to consider
bulk gauge fields, one should incorporate some mechanism, such as
non-trivial compactifications, in order to remove such adjoint
scalars.

\section{Effects of the NG boson $\phi$}
\setcounter{equation}{0}

In this section, we investigate the effects of the Nambe-Goldstone
field $\phi(x)$ on our four-dimensional world using the action
obtained in the previous section. We take the dimension of our 
brane $d=4$ and denote the number of extra 
dimensions $(D-d)\equiv \delta$. It is also simply assumed that as in
the usual case, the gravity (and sometimes the standard gauge fields,
too) live in the bulk and the other matter fields are all confined in
the four-dimensional brane. For simplicity, we use a dimension one
parameter $f$ for the brane tension in the following:
\begin{eqnarray}
  \tau &\equiv& f^4/4\pi^2.
\end{eqnarray}

Strictly speaking, when we take the bulk gravity turned on, the NG
bosons $\phi^m$ are absorbed by the zero modes of the off-diagonal
components of the bulk metric, $B^m_\mu$, and the latter become
massive. Their masses, however, become of the 
order $\sim\sqrt{\tau}/M_{\rm pl}$~\cite{tev-ph} ($M_{\rm pl}$ is the
Planck mass in four dimensions), and very small compared to the energy
scales which we will consider. Therefore, we need not take into
account this Higgs effect thanks to the equivalence 
theorem~\cite{equiv}, and will treat $\phi$ (and graviton zero modes)
to be massless fields.

\subsection{Suppressions of KK mode couplings}

First, we briefly review the results of Ref.~\cite{BKNY} about the
couplings (\ref{normal}) between the brane fields and the bulk
ones. As we will see below, the couplings can exponentially reduce the
contributions of higher KK modes and potentially make their effects
invisible in our four-dimensional world.

The form of the interaction term (\ref{normal}) now implies that the
effective coupling $g_n$ of the level $n$ KK mode to four-dimensional
fields is actually suppressed exponentially:
\begin{eqnarray}
  g_n &\equiv& g\cdot 
  e^{-\frac{1}{2}\left(\frac{n}{R}\right)^2 \frac{M^2}{f^4}}\,.
\end{eqnarray}
The origin of this suppression is a recoil effect of the brane. This
is easily seen if we note that the effective couplings $g_n$ can be
written as
\begin{eqnarray}
  g_n &=& g\cdot\langle 0|\, e^{2\pi i\frac{n}{R}\frac{\phi(x)}{f^2}}
  \, |0 \rangle,
\end{eqnarray}
by using the perturbative vacuum $|0 \rangle$ of the NG 
bosons $\phi$. Remembering that $\phi(x)$ stand for the transverse
coordinate, the operator $e^{i\frac{n}{R}\phi(x)}$ is just like the
vertex operator in the string theory and gives transverse 
momentum $n/R$ to the brane around the point $x$. 
Hence, $\,e^{2\pi i\frac{n}{R}\frac{\phi(x)}{f^2}} |0\rangle$
represents the recoiled state of $\phi$ by the absorption (emission)
of the level $n$ KK mode carrying transverse 
momentum $n/R$ ($-n/R$). Thus the amplitude $\,\langle0|\, 
e^{2\pi i\frac{n}{R}\frac{\phi(x)}{f^2}}\, |0\rangle\,$ can be viewed
as a probability amplitude of containing the original 
state $|0\rangle$ in the recoiled 
state $e^{2\pi i\frac{n}{R}\frac{\phi(x)}{f^2}} |0\rangle$. As is
clear from this view, the suppressions become stronger for higher KK
modes since larger deformation of the brane occurs, and on the other
hand, in the case of {\it stiff} \/ brane possessing large $f$, the
suppression is weak.

Let us discuss a phenomenological consequence of this new suppression
factor. Consider a tree-level amplitude for the two charged-particle
scattering in the brane caused by the exchange of KK gauge bosons
which live in the bulk. This leads to a correction to the effective
four-Fermi coupling constant $G_F$. The summation over all the KK
modes in the amplitude becomes
\begin{eqnarray}
  \sum_n\, g^2_n\, \langle A^{(n)}_\mu A^{(-n)}_\nu \rangle &\sim& 
  \sum_n g_n^2\, \frac{1}{M_W^2+n^2/R^2},
  \label{An}
\end{eqnarray}
where we have assumed the momentum transfer is small compared 
with $M_W$ (and then $R^{-1}$). If the couplings of KK modes were
universal, $^\forall g_n =g$, this amplitude diverges when the
dimensions transverse to the brane are greater than one. This is
something wrong because it is merely a tree-level amplitude and the
divergence cannot be renormalized by any means. In the recent analyses
on the experimental implications of such KK modes contributions, the
sum has simply been cut off at the fundamental scale $M$. However, we
see above that such divergences are automatically cured as it should
by properly taking into account the brane fluctuations (the
fluctuations of $\phi$).\footnote{Similar suppression factors are also
  discussed in different frameworks~\cite{suppress}.}

For a numerical estimation, we simply consider the $\delta=1$ case,
but it is straightforward to include more numbers of extra
dimensions. The correction (\ref{An}) is dominated by the first mode
contributions (with $n=\pm 1$) since the higher modes are further
suppressed by the presence of exponential factors, so that the KK mode
correction $\Delta G_F$ to the four-Fermi coupling constant can be
estimated as
\begin{eqnarray}
  \frac{\Delta G_F}{G_F} &\sim& 2\frac{g^2_{n{=}1}}{g^2}
  \frac{M_W^2}{M_W^2+1/R^2}\,.
  \label{corr}
\end{eqnarray}
Since the standard model prediction precisely agrees with experiments,
the KK mode correction (\ref{corr}) must be small, 
say ${\raise2pt\hbox{$<$}}\kern-9pt\lower4pt\hbox{$\sim$}\,
O(10^{-2})\,$~\cite{PDG}. For definiteness, we consider $R^{-1}$ in
the range $M_W < R^{-1}\, (< M)$, then the 
constraint $\Delta G_F/G_F < 10^{-2}$ reads
\begin{eqnarray}
  2M_W^2 R^2\cdot e^{-\frac{M^2}{R^2f^4}} \,<\, 10^{-2}\,.
  \label{const}
\end{eqnarray}
This constraint gives a weak upper bound to the brane tension; 
$\,f\; {\raise2pt\hbox{$<$}}\kern-9pt\lower4pt\hbox{$\sim$}\; O(1)$
TeV, depending on the value of $M$. Clearly, 
since $\exp[-(M^2/R^2f^4)] < 1$, no constraint appears if $R^{-1}$ is
larger than $10\sqrt{2}M_W\sim 1.1$ TeV\@. Even for $R^{-1}$ less than
this, the constraint (\ref{const}) can be easily satisfied for any
value of $M$ provided that $f$ is chosen suitably small. Therefore, if
the brane fluctuation is taken into account, the constraints on the
extra dimensions ($R, M$) so far obtained can be substantially
loosened and sometimes disappear. In the case of $\delta \geq 2$,
since the exponential damping factor appears for each extra dimension,
it is clear that the suppression factor becomes $(g_n/g)^{2\delta}$
and then, the constraint of $f$ is further weaker.

Finally, it should be noted that the exponential suppressions
discussed above also work in the couplings of the KK gravitons to the
matters on the brane. This fact will have important effects in later
analyses.

\subsection{The fifth force}

In the previous section, we have shown that the contributions from all 
KK excited modes are suppressed if the brane tension $f$ takes a 
moderately small value. Then, we unfortunately could not find
signatures of the presence of large extra dimensions unless there is a
lower bound for $f$. In section 3, we have seen two types of
interaction terms for the NG field $\phi(x)$ involving the 
coupling $f$. One is the exponential couplings between the brane
fields and the bulk ones whose effect was discussed before. It gave
only upper bounds of $f$. Expanding the exponential interaction, we
also have the $n$ NG boson coupling whose strength in general takes
the form $\sim (1/Rf^2)^n\exp[-M^2/R^2f^4]$. This factor takes 
a {\it maximum} \/ value at $f \sim \sqrt{M/R}$, which is independent
of $f$, and so cannot be used to give a lower bound for $f$ either. On
the other hand, another type of couplings in the 
lagrangian (\ref{SNG}), coming from the expansion of the vierbein, has
strength proportional to $f^{-2}$. As the brane tension becomes
smaller, its effects become larger and may easily be detected. In the
following two sections, we will calculate various phenomena which
involve the latter coupling, and show that relatively severe lower
bounds can be obtained.

First, we discuss a constraint from the so-called ``fifth force''. So
far, various types of new long range forces mediated by hypothesized
very light particles have been suggested in many theoretical
frameworks~\cite{5thforce}. We here consider the strength of a long
range force which arises as a consequence of exchanges of the NG
scalar $\phi$, which could be potentially observable. In the
macroscopic range, the dominant force working between neutral systems
is gravity. The Newton's universal law of gravitation (the
inverse-square distance dependence) has been experimentally tested up
to 1 cm range, and new forces which are comparable to gravity have
been excluded to this range~\cite{gravity}. This fact restricts the
potential forms of new forces and imposes the bounds for new couplings
which determine the potentials. In the present model, this just
results in a lower bound for $f$. As seen from the effective
lagrangian on the brane, the NG mediated force between two
distinguishable standard fermions with masses $m$ and $m'$ is
calculated from the one-loop diagram in the leading order 
of $f^{-1}$ (Fig.~1). 
\begin{figure}[htbp]
  \begin{center}
    \leavevmode
    \epsfxsize=5cm \ \epsfbox{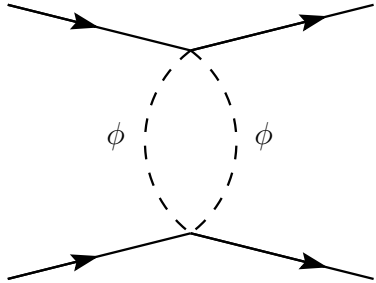}
    \put(-103,53){$\phi$}
    \put(-47,53){$\phi$}
  \end{center}
  \caption{The lowest-order contribution to the NG boson
    mediated force.}
\end{figure}
Note that since a single $\phi$ does not have tree-level couplings to
the standard fermions, this two-particle exchange is the leading
contribution. This is because in the action on the brane, there is 
an $SO(\delta)$ internal symmetry under which $\phi$ transform as a
vector. So, unlike the case of Yukawa force mediated by massless
scalars, the above new force at a distance $r$ is expected to behave
as $1/r^n\, (n>2)$. We calculate the above amplitude in the
non-relativistic limit and extract the two-body potential $V(r)$ from
it. When we now consider a loop diagram and a higher-order correction
to gravity, a convenient way to compute the two-body potential is the
dispersion theoretical method~\cite{dispersion}. This method gives a
following relation between the potential $V(r)$ and invariant
scattering amplitude ${\cal M}$ obtained from Feynman diagrams,
\begin{eqnarray}
  V(r) &=& \frac{i}{8\pi^2 r} \int_0^\infty dt\, {\cal M}_t 
  \exp (-\sqrt{t}\, r),
  \label{disp}
\end{eqnarray}
where $t$ denotes the momentum transfer (that is now used beyond the
physical region) and ${\cal M}_t$ is the discontinuity of amplitude
across the branch cut on the real $t$ axis; ${\cal M}_t \equiv 2i\,
{\rm Im}\, {\cal M}(t +i\epsilon)|_{\epsilon\to +0}$.
Therefore, in order to obtain the expressions for long range
potentials we have only to calculate the $t$-channel discontinuity and
perform a Laplace transformation. The discontinuity of amplitudes can
be evaluated by a simple prescription resulting in the replacement of
each propagator by its discontinuity (delta function) multiplied by a
theta function to insure a positive energy~\cite{dispersion};
\begin{eqnarray}
  \frac{1}{m^2-k^2} &\longrightarrow& 2\pi i\, \delta(m^2-k^2)
  \theta(k^0).
\end{eqnarray}
For the above diagram, we obtain for ${\cal M}_t$ in the
non-relativistic limit neglecting the higher-order contributions of
three-momenta,
\begin{eqnarray}
  {\cal M}_t &=& \frac{\pi^3i\,\delta\, mm'}{80 f^8}\, t^2, 
\end{eqnarray}
where $\delta$ is the number of extra dimensions. We here have used a
formula
\begin{eqnarray}
  \int d^4k\, \delta\bigl(k^2\bigr)\,
  \delta\bigl((k-q)^2\bigr)\, k^\mu k^\nu k^\rho k^\sigma \,=\,
  \frac{\pi}{10}\left(q^\mu q^\nu q^\rho q^\sigma -\frac{1}{8} 
    g^{\left(\mu\nu\right.} q^\rho q^{\left.\sigma\right)} q^2
    +\frac{1}{48} g^{\left(\mu\nu\right.}\! 
    g^{\left.\rho\sigma\right)} q^4 \right),
\end{eqnarray}
where two totally symmetric sums are defined as
\begin{eqnarray}
  g^{\left(\mu\nu\right.} q^\rho q^{\left.\sigma\right)} &=&
  g^{\mu\nu} q^\rho q^\sigma +g^{\mu\rho} q^\nu q^\sigma
  +g^{\mu\sigma} q^\nu q^\rho +g^{\nu\rho} q^\mu q^\sigma
  +g^{\nu\sigma} q^\mu q^\rho +g^{\rho\sigma} q^\mu q^\nu, \\
  g^{\left(\mu\nu\right.}\! g^{\left.\rho\sigma\right)} &=& g^{\mu\nu}
  g^{\rho\sigma} +g^{\mu\rho} g^{\nu\sigma} +g^{\mu\sigma}
  g^{\nu\rho}.
\end{eqnarray}
As a result, we see that this force has a potential
\begin{eqnarray}
  V(r) &=& -\frac{3\pi\delta}{8 f^8} \frac{mm'}{r^7},
  \label{force}
\end{eqnarray}
implying an attractive long range force. In Appendix B, we show
another derivation of this result by using more conventional method.

Before performing numerical comparisons with the experiments, we here
give several comments on other possible contributions besides the
above potential. First, there is a contribution from the massive KK
gravitons. The deviations from the Newton's law by the KK graviton
mediated force becomes Yukawa-type potentials~\cite{deviation} and can
be negligible in the macroscopic region. On the other hand, when the
number of extra dimensions is two this force can be as strong as the
four-dimensional gravity at sub-millimeter range for the fundamental
scale $M \simeq 1$ TeV\@. (In that case, moreover, the KK graviton
mediated amplitude has a logarithmic dependence on the brane 
tension $f$ and cannot be suppressed enough for any small values 
of $f$.) \ However, several astrophysical constraints have already
excluded the region for $M$, at least, up to $O(10)$ 
TeV~\cite{astroph} and in addition, the KK graviton mediated force has
a strong dependence of $M^{-4}$. So, we can neglect this KK graviton
contributions even at the level of the proposed gravitational
experiments. In the case of $\delta \geq 3$, the KK graviton force is
much weaker than the gravitation due to the very short Compton
wavelengths of the KK gravitons. There are also finite temperature
corrections which, in the long distance range, could become comparable
to the zero-temperature contribution calculated above. So the
potential $V(r)$ may receive the correction of the factor of 
order $O(1)$. However, since the potential has a large power
dependence on the brane tension $f$, such an $O(1)$ correction is
little relevant in evaluating the constraints for $f$ numerically. We
therefore neglect the finite temperature corrections.

There are some types of experiments which can test deviations from the
inverse-square law of the Newtonian gravity. The torsion balance
experiments give the most stringent limits on the deviations up to
sub-centimeter range and have excluded the presence of new forces
whose strengths are comparable to or stronger than the 
gravity~\cite{torsion}. Furthermore, the electromagnetic Casimir force 
have recently been measured very precisely and also can impose the
constraints on new forces in the range below $10^{-4}$ 
m~\cite{casimir}. By comparing our hypothetical force (\ref{force})
with the results of those experiments, we obtain a constraint for $f$
\begin{eqnarray}
  f &>& O(0.1) {\rm ~GeV}.
  \label{nforce}
\end{eqnarray}
Because of the steep slope of $V(r)$ proportional to $r^{-7}$, this
lower bound mainly comes from the Casimir force experiments. The
proposed classical gravity experiments will further improve this bound
by a factor of $O(1)$.

The bound (\ref{nforce}) is a very weak one. But, it certainly shows
that there exists a lower bound for the brane tension. 

\subsection{Energy loss in stars}

The fifth force constraint in the previous section is too weak to give
an enough bound for the brane tension $f$. However, there are other
processes which actually can lead severer bounds for $f$. Since $\phi$
does not have any gauge charges on the brane, its properties can only
be constrained by the missing energy arguments in the collider
experiments and in astrophysics. There, one usually assumes that the
energy loss occurs via the standard mechanism in form of photons
and/or neutrinos. For example, if there are novel low-mass particles
having weak interactions with matters, they can freely escape from the
interior of stars and carry away their energy. Then it will change the
course of the stellar evolution that would be expected otherwise. With
this observation in various astronomical observables such as the Sun,
horizontal branch stars, white dwarfs, supernovas, we can derive
various types of bounds on the coupling strengths in the models under
consideration~\cite{raffelt}. In this section, we discuss constraints
on the brane tension $f$ required from the cooling process of the
neutron star in supernova explosions. That is, if there were novel
particles which freely stream out of the nascent neutron star, it may
rather affect the observed data of neutrino burst. This will lead to
the most stringent bounds from astrophysics. The other collider
signatures of energy loss carried away by the NG field $\phi$ are also
important and will be discussed elsewhere.

We can obtain a rough estimation for this supernova bound by comparing
naively with the axion case. Though this estimation is very rough, we
can see that the astrophysical requirements actually impose severer
bounds for the brane tension. The axion coupling to the ordinary
matter is proportional to $1/f_a$, where $f_a$ is the axion decay
constant. The cross section of axion emission from the interior core
is proportional to $1/f_a^2$. The astrophysical constraints from the
supernova observation say that $f_a$ should be greater 
than $10^{10}$ GeV~\cite{raffelt}. On the other hand, a typical
coupling of the NG boson $\phi$ to matters are given 
by $\sim T_{SN}^3/f^4$, where $T_{SN}$ is the supernova 
temperature (available energy in the supernova medium). If we
naively convert the axion bound to the present case, then we obtain
\begin{eqnarray}
  \frac{T_{SN}^3}{f^4} \;\,<\,\; \frac{1}{10^{10} {\rm ~GeV}}\;
  &\longrightarrow& \; f \,\;
  {\raise2pt\hbox{$>$}}\kern-9pt\lower4pt\hbox{$\sim$}\;\,
  10^{1.5} {\rm ~GeV},
  \label{rough}
\end{eqnarray}
for the supernova temperature $T_{SN}\sim O(10)$ MeV\@. This shows
that, as anticipated, the star energy loss argument certainly gives a
severer bound for the brane tension than that from the fifth force
constraint.

We now perform a more detailed analysis of the energy emission rate
from the supernova core. There are several channels which result in
energy losses in the supernova medium. We expect that the most
important contribution to this is the nucleon-nucleon bremsstrahlung
process like in the axion case~\cite{axion} because the supernova
temperature is near the pion mass and large suppressions with the pion
mass do not occur. However, this process is rather involved and there
are many types of uncertainties and unknown factors, for example, with
reference to the effective pion description in the supernova
medium. Instead, in this paper we consider a more ``clean'' event,
i.e., the electron-positron pair annihilation process. This
contribution is surely subdominant but can be as large as that of the
nucleon bremsstrahlung when we adopt a higher value of temperature in
the supernova. However, we should bear in mind that more stringent
bounds may be obtained by accurately taking into account the dominant
process mentioned above.

Since the energy scale considered is well below the fundamental 
scale $M$ and the tension $f$, there are two diagrams contributing to
the energy losses from the electron-positron annihilation in the
leading order in the couplings $G_N$ and $g_i$ ($G_N$ is the Newton
constant and $g_i$ are the standard gauge couplings). They are
annihilations to (a) the KK gravitons $g_{\mu\nu}^{\,(n)}$ and (b) the
NG bosons $\phi$ (Fig.~2).
\begin{figure}[htbp]
  \begin{center}
    \leavevmode
    \epsfxsize=3.5cm \ \epsfbox{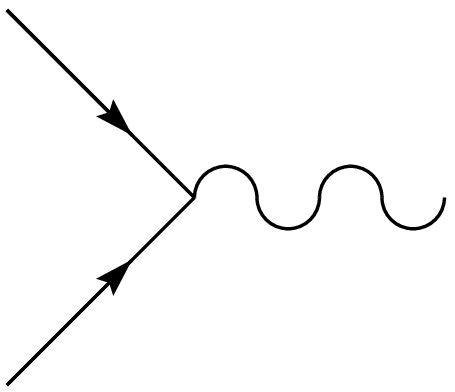}
    \put(-115,84){$e^+$}
    \put(-115,0){$e^-$}
    \put(8,42){$g_{\mu\nu}^{\,(n)}$}
    \put(-137,42){$\displaystyle{\sum_n}$}
    \put(-60,-24){(a)}
    \hspace*{3.5cm}
    \epsfxsize=3.5cm \ \epsfbox{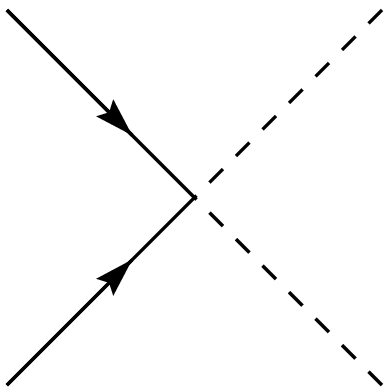}
    \put(-115,95){$e^+$}
    \put(-115,0){$e^-$}
    \put(4,97){$\phi$}
    \put(4,0){$\phi$}
    \put(-57,-24){(b)}
  \end{center}
  \caption{The energy loss processes in the supernova medium by the
    electron-positron annihilation: (a) KK graviton production and (b) 
    NG boson pair production.}
\end{figure}
We estimate the initial-spin averaged cross section for each case (a)
and (b). The explicit calculation is given in Appendix C\@. For the
real KK gravitons production process, the cross section is obtained by
only attaching an exponential suppression factor discussed before to
the usual result;\footnote{If we sum up the inclusive 
  process $e^++e^- \to g_{\mu\nu}^{(n)}+$ (any numbers of $\phi$) in
  place of the exclusive process (a), the cross section $\sigma_a$ may
  be enhanced by a factor of $\exp(as^2/f^4)$ with an $O(1)$
  coefficient $a$, which is, however, not large enough to cancel the
  suppression factor $\exp(-sM^2/f^4)$.}
\begin{eqnarray}
  \sigma_a &=& \frac{\pi^{1-\frac{\delta}{2}}}{2^{\delta+3}
    \Gamma(\frac{\delta}{2})}\frac{1}{M^{\delta+2}}\,
  s^\frac{\delta}{2}\, e^{-s \frac{M^2}{f^4}}, \\
  \sigma_b &=& \frac{\delta\pi^3}{1920}\frac{1}{f^8}\, s^3,
\end{eqnarray}
where $s$ is the center of mass energy in the electron-positron
system. We have used a 
relation $G_N^{-1}=8\pi V_\delta M^{\delta+2}$ where $V_\delta$ is the
volume of the $\delta$-torus ($=(2\pi R)^\delta$). We have also
neglected the electron mass for $m_e \ll T_{SN}$ in the core. It
should be noted that the $f$-dependences of the above two cross
sections are very different from each other. Typical behaviors of the
cross sections are shown in Fig.~3.
\begin{figure}[htbp]
  \begin{center}
    \leavevmode
    \epsfxsize=8cm \ \epsfbox{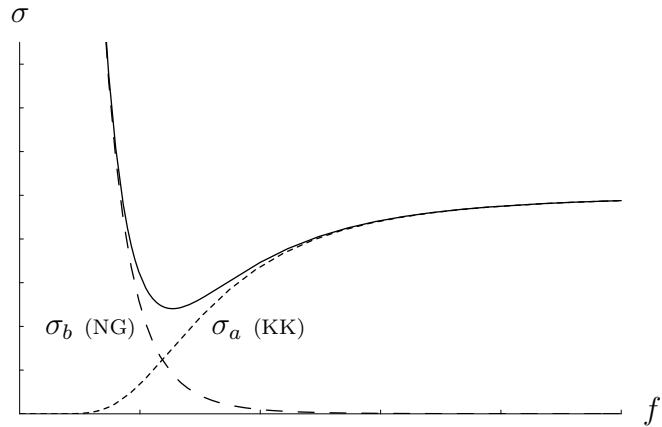}
    \put(8,10){$f$}
    \put(-231,160){$\sigma$}
    \put(-155,42){$\sigma_a$ {\scriptsize (KK)}}
    \put(-218,42){$\sigma_b$ {\scriptsize (NG)}}
    \caption{The typical behaviors of the cross sections against the
      brane tension $f$. The dotted and dashed lines denote $\sigma_a$
      and $\sigma_b$, respectively. The solid line is the total
      cross section.} 
  \end{center}
\end{figure}
It can be seen from this figure that the total cross section (the
solid line in Fig.~3) becomes large for both small and large values
of the brane tension $f$. The cross section for a small value of $f$
is governed by the NG boson production process while for a large
value, the KK gravitons production rate dominates the total cross
section. In either way, it turns out that we cannot have arbitrarily
desired values of brane tension and it will be severely restricted by
phenomenological requirements.

If the present novel particles interact strongly with ordinary
matters, they are scattered or absorbed again in the medium after
productions in the core. Then they cannot freely escape to the outside
and will be radiated from a sphere. The flux from this black-body
surface emission for the neutron star becomes smaller as novel
particles interact more strongly (the radius of sphere becomes
larger). In that case, the region beyond some strong coupling cannot
be excluded from the cooling argument like in the axion 
case~\cite{axion-trap}. For the KK gravitons, the energy reabsorption
is, however, highly suppressed. This can be intuitively understood
from the fact that the brane occupies a very tiny (almost zero) region
in the whole bulk. Once the KK gravitons escape to the bulk spacetime
from the brane, they will never return, at least, over the age of the
universe~\cite{tev-ph}. After all, the KK gravitons can be considered
to freely stream out of the core. On the other hand, since $\phi(x)$
is a field on our four-dimensional brane and does not escape to the
extra dimensions, its effect may become less important than neutrinos
for the strong coupling (small tension) region. For this, we estimate
the mean free path $L$ for $\phi$ in the core, which is roughly given
by
\begin{eqnarray}
  L &\sim& \frac{f^8}{T_{\rm SN}^6 n_e},
\end{eqnarray}
where $n_e$ is the number density of electron in the 
core, $n_e \sim 10^{-3}$ GeV$^3$. We can see that $L$ exceeds the
radius of the neutron star $\sim$ 10 km for
$f\>{\raise2pt\hbox{$>$}}\kern-9pt\lower4pt\hbox{$\sim$}\> O(10)$
GeV (see, (\ref{rough}) or (\ref{fbound})). Then, $\phi$ is emitted
from the entire volume of the star and we can surely discuss the lower
bounds of brane tension $f$. Note that if the couplings were strong
and the mean free path $L$ were smaller than the star radius, other
astrophysical constraints and/or too many experimental signals in the
neutrino detectors~\cite{axion-exp} would reject such strong coupling
regions.

From the calculated cross sections, we extract the production rates in
order to compare them with the observations and to have definite 
conclusions about the allowed range of the parameters. However, in
the hot dense medium as in the supernova, there are some uncertainties
such as many-body effects, which have not been fully understood. We
expect that those collective effects in the medium amount to change a
factor of $O(1)$ in the production rate. This is actually the case for
the axion emission from the supernova~\cite{raffelt}. Moreover, as we
will see below, the production rates have large power-dependence on
parameters for which we now would like to have limits. So, even if
there is an ambiguity even of factor of 10, the final results for
these parameters are not so affected.

Now, let us estimate the bounds numerically. The theory of type II
supernova explosions says that in the explosion, almost all of the
gravitational binding energy of a nascent neutron star is released in
form of neutrinos within a few second. The expected neutrino flux and
the duration of signal calculated from this picture is in good
agreement with the observations for the supernova 1987A in the
Kamiokande~II~\cite{kamioka} and IMB~\cite{IMB} detectors. This
agreement implies a conservative requirement that the energy loss rate
from other than neutrinos should not exceed that from neutrinos in the
standard picture~\cite{raffelt}. Otherwise the measured duration of
the neutrino signal would be shorter than the observed
ones. Numerically, this constraint reads
\begin{eqnarray}
  Q &{\raise2pt\hbox{$<$}}\kern-9pt\lower4pt\hbox{$\sim$}& 10^{36} 
  {\rm ~GeV/cm}^3\, {\rm sec}.
  \label{SNbound}
\end{eqnarray}
Here, $Q$ is the energy loss rate per unit time and unit volume (the
volume emissivity) defined as follows;
\begin{eqnarray}
  Q_x &\equiv& \int \frac{d^3 k_1}{(2\pi)^3 2 E_1} \int 
  \frac{d^3 k_2}{(2\pi)^3 2 E_2}\, 2f_1\cdot 2f_2\cdot (E_1+E_2)\cdot
  2 s\, \sigma_x,
\end{eqnarray}
where $x$ is the process label, $a$ or $b$, and $f_i$ is the
Fermi-Dirac distribution function; $f_i=1/(\exp(E_i/T-\mu/T)+1)$ (a
factor of 2 denotes the spin degrees of freedom). For $\mu$ in the
distribution functions, we use the chemical potential for electron
(positron) in the supernova core. Since the electron is considered as
a highly degenerate relativistic particle in the core, $\mu$ is given 
by $\simeq (3\pi^2 n_e)^{1/3}$ where $n_e$ is the number density of
electron, $n_e \sim 1.4\times 10^{-3}$ GeV$^3$. We calculate $Q$ for
each process (a) and (b), and the results become
\begin{eqnarray}
  Q_a &=& \frac{1}{4\pi^{\frac{\delta}{2}+3}\Gamma(\frac{\delta}{2})}
  \frac{T_{SN}^{\,\delta+7}}{M^{\delta+2}}\, F_\delta(T_{SN}), \\
  Q_b &=& \frac{\delta}{75\pi}\frac{T_{SN}^{13}}{f^8}\, I(T_{SN})\,.
\end{eqnarray}
The dimensionless functions $F_\delta(T)$ and $I(T)$ are given by
\begin{eqnarray}
  F_\delta(T) &\equiv& \int_0^\infty \int_0^\infty dxdy
  \frac{(x+y)(xy)^{\,\delta/2+2}}{(e^{x-(\mu/T)}+1)(e^{y+(\mu/T)}+1)}\,
  \Gamma_\delta (4xy T^2 M^2/f^4), \\
  I(T) &\equiv& \int_0^\infty \int_0^\infty dxdy
  \frac{(x+y)(xy)^5}{(e^{x-(\mu/T)}+1) (e^{y+(\mu/T)}+1)},
\end{eqnarray}
where $\Gamma_\delta (z)$ is defined as
\begin{eqnarray}
  \Gamma_\delta (z) &\equiv& \int_0^1 dt\, t^{\,\delta/2+1}\, e^{-zt}.
\end{eqnarray}
Clearly, the relation $\Gamma_{\delta+2n}(z)=(-d/dz)^n
\Gamma_\delta(z)$ holds and we have
\begin{eqnarray}
  \Gamma_\delta(z) &=& \left(-\frac{d}{dz}\right)^{n+1} \cases{
    \displaystyle\frac{1-e^{-z}}{z} & ~for $\delta=2n$ \cr
    \noalign{\vskip1ex}
    \displaystyle\sqrt{\frac{\pi}{z}}\, {\rm Erf}\, (\sqrt{z}) & ~for
    $\delta=2n-1$}
\end{eqnarray}
where Erf$\,(z)$ is the error function,
\begin{eqnarray}
  {\rm Erf}\,(z) &\equiv& \frac{2}{\sqrt{\pi}} \int_0^z\, e^{-t^2} dt.
\end{eqnarray}

From the above results, we can obtain two types of bound concerning
with extra dimensions in the present model. This is mainly because the
two volume emission rates have very different parameter dependences as
mentioned before (see, Fig.~3). First, we discuss the lower bounds
for the brane tension $f$. As can be seen from Fig.~3, for a small
value of $f$, the dominant contribution to the energy loss rate is
that from the NG bosons production. We apply the supernova 
constraint (\ref{SNbound}) for the volume emissivity to $Q_b$ and then
obtain lower bounds (in the case of $\delta=1$);
\begin{eqnarray}
  f &>& \; 8\> {\rm ~GeV} \qquad (T_{SN} = 20 {\rm ~MeV}), \\
  f &>& 23 {\rm ~GeV} \qquad (T_{SN} = 30 {\rm ~MeV}), \\
  &\vdots&  \nonumber \\
  f &>& 122 {\rm ~GeV} \quad\;\; (T_{SN} = 70 {\rm ~MeV}).
  \label{fbound}
\end{eqnarray}
Here we have adopted a rather wide range of the supernova temperature
and displayed corresponding bounds for each case. Note that if there
are $\delta$ extra dimensions, each value of the lower bound is
multiplied by $\delta^{1/8}$. It is interesting that the above bounds
have no dependence on the value of the fundamental scale $M$ since the
cross section for NG bosons production is independent of it. The above
limits are indeed $10^{2-3}$ times stronger than that from the fifth
force constraint. In other words, with these bounds the long range
force associated with $\phi$ is outside the reach of the proposed
gravity experiments, and it also does not modify the effects of
gravity in stars. In this way, the brane tension can be severely
restricted from astrophysical arguments and the KK mode couplings
cannot be arbitrarily suppressed. This is a welcome result for we
could see signatures of the existence of extra dimensions in the near
future experiments.

More interestingly, we can also get the lower bounds for the
fundamental scale $M$. As depicted in Fig.~3, there is an absolute
minimum of the total cross section against the brane tension $f$. That
is, the star energy is necessarily carried away to a certain extent in
form of the KK gravitons and NG bosons. At the minimum, the value of
$f$ and then the minimum value of the energy loss rate are determined
by a given value of $M$. Therefore, as a conservative bound for $M$,
we require that the total energy loss rate at the minimum must not
violate the supernova condition (\ref{SNbound}). It is clear that
the value of $f$ minimizing the emission rate satisfies the above
obtained lower bounds, and moreover the resultant bounds for $M$
become weaker than that naively estimated in the TeV-scale quantum
gravity scenario~\cite{astroph}, i.e., without the coupling
suppression by the brane fluctuations taken into account. The cross
section for the KK graviton emissions (and hence $Q_a$) rather depends
on the number of extra dimensions. It can be easily seen that 
the $\delta=2$ case gives the maximum cross section and hence we
obtain the most restrictive bound. For $\delta>2$, the cross sections
are considerably reduced with increasing power of $M^{-1}$, and we
only obtain very weak lower bounds for $M$. The fundamental scale
dependences of the energy loss rate at the minimum are shown in 
Fig.~4 for the $\delta=2$ case. Remarkably, the fundamental scale $M$
is strongly restricted even when our four-dimensional brane is
relatively soft; e.g., $M > O(1)$ TeV for $T_{SN}=30$ MeV\@.
\begin{figure}[htbp]
  \begin{center}
    \epsfxsize=9.5cm \ \epsfbox{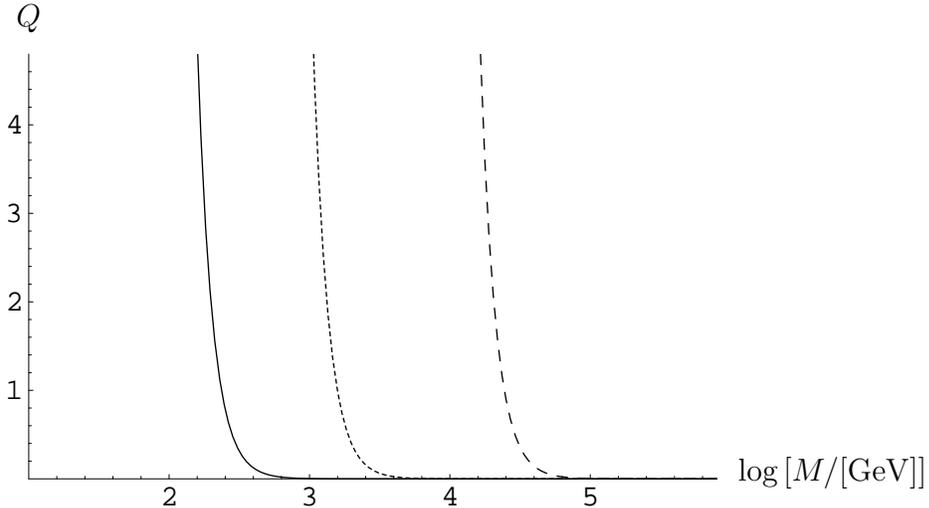}
    \put(8,14){$\log\,[M/[{\rm GeV}]]$}
    \put(-265,186){$Q$}
    \caption{The volume emissivity $Q$ in the supernova core for
      $\delta=2$ case. The solid, dotted and dashed line correspond to
      $T_{SN}=20, 30, 70$ MeV, respectively. ($Q$ is denoted in units
      of $10^{36}$ GeV/cm$^3$ sec.)}  
  \end{center}
\end{figure}
Notice that we may have more severe bounds for parameters if we
consider the dominant process of nucleon bremsstrahlung in the
supernova core as noted before, and moreover other experimentally
observable processes may give considerable impacts on the model
parameters. Anyway, we cannot have arbitrarily small values for the
brane tension and therefore do not miss the possibility of finding
signatures of the extra dimensions.

\section{Summary}

In this paper, we have investigated the possibility where our
four-dimensional world is embedded as a brane in higher dimensional
spacetime. In such a situation, the translational symmetry transverse
to the brane is spontaneously broken by the presence of brane
itself. As a consequence, the massless NG fields appear which denote
the position of brane in the higher dimensional bulk. For such a
setup, the effective action for the four-dimensional fields is
described by using the induced metric and vierbein on the brane. These
induced quantities can be given in terms of the bulk vielbein and the
NG variables. We have found the explicit expression for the induced
vielbein and written down the effective action for fermion
fields confined on our brane. Given the interactions, we can discuss
phenomenological implications of the brane effective action,
especially, several interesting effects of the NG fields.

First, we discussed the KK mode couplings to the brane. We found 
in~\cite{BKNY} that higher KK mode contributions are exponentially
suppressed with a suitably small tension of our world, i.e., for a
relatively soft brane. However, some other phenomenological
observations can impose limits on this softness. We have shown in
practice that severe lower bounds for the brane tension can be
obtained by considering, for example, the null observations of the
fifth force, the missing energy process in the supernova explosions
such as SN1987A, and so on. The existence of these lower bounds
implies that the KK mode contributions cannot be arbitrarily
suppressed and could be observable in the near future experiments as
has been discussed so far in many articles. We have also been able to
calculate a lower bound for the fundamental scale of the effective
theory of our brane. More promising and severe bounds for the physics
of extra dimensions (the fundamental scale, the compactification
radius, etc.) would be obtained by studying their signatures in the
collider experiments and other astrophysical and cosmological
requirements. Hence, it will be an interesting and challenging subject
to find evidences for the existence of extra spacetime dimensions in
form of the Kaluza-Klein modes and/or the NG bosons appearing on our
four-dimensional world.

\subsection*{Acknowledgements}
We would like to thank M.~Bando, K.~Hashimoto, T.~Hirayama,
T.~Noguchi, N.~Sugiura for valuable discussions and 
comments. T.~K.\ and K.~Y.\ are supported in part by the Grants-in-Aid
for Scientific Research No.~10640261 and the Grant-in-Aid No.~9161,
respectively, from the Ministry of Education, Science, Sports and
Culture, Japan.

\newpage

\appendix
\section{Induced vielbein}
\setcounter{equation}{0}

In this appendix, we derive an explicit form of the induced vielbein
on the $d$-dimensional brane. It is written in terms of 
the $D$-dimension bulk vielbein and the $(D-d)$ NG variables denoting
the position of brane in the bulk.

The $d$-dimensional brane breaks the bulk Lorentz group $G=SO(1,D-1)$
down to $H=SO(1,d-1)\times SO(D-d)$. Consider the coset-space 
variable $\xi(\theta)\in G/H$,
\begin{eqnarray}
  \xi(\theta(x)) &=& \exp(i\theta^\alpha_{\ a}(x)\eta_{\alpha\beta}
  J^{(\beta a)}),
\end{eqnarray}
with broken generators $J^{(\alpha a)}$. The 
fields $\theta^\alpha_{\ a}(x)$ is the NG boson associated with the
spontaneous breaking of $G$ down to $H$. Act the bulk Lorentz
transformation $g\in G$ to the element $\xi(\theta(x))$ from the left,
then the resultant $g\xi(\theta(x))$ is still an element of $G$ and
can be decomposed uniquely into the form:
\begin{eqnarray}
  g\xi(\theta(x)) &=& \xi(\theta'(x))h(g,\theta(x)), \qquad ^\exists
  h(g,\theta(x))\in H.
  \label{defh}
\end{eqnarray}
Thus, the bulk Lorentz transformation $g\in G$ for the NG 
variable $\theta(x)$ is defined by
\begin{eqnarray}
  && \xi(\theta(x)) \;\; \stackrel{g}{\longrightarrow} \;\;
  \xi(\theta'(x)) \;=\; g\xi(\theta(x))h^{-1}(g,\theta(x)).
\end{eqnarray}
An important property of $\xi(\theta(x))$ is that the $G$
transformation $g$ is converted through $\xi(\theta(x))$ into the
corresponding $H$ transformation $h(g,\theta(x))$. 


Up to here everything is the standard story of the nonlinear
realization theory~\cite{BKY}. We wish to define the 
vielbein $e^\alpha_{\ \mu}$ induced on the brane from the bulk
vielbein $E^A_{\ M}$. The necessary conditions for the desired induced
vielbein $e^\alpha_{\ \mu}$ to satisfy are that (i) $e^\alpha_{\ \mu}$
for each $\mu$ transforms as a $d$-dimensional vector 
under $SO(1,d-1)$. (ii) the metric given by $e^\alpha_{\ \mu}$ should
coincides with the induced metric;
\begin{eqnarray}
  &&g_{\mu\nu} \,=\, \eta_{\alpha\beta}e^\alpha_{\ \mu}e^\beta_{\ \nu} 
  \,=\, G_{MN}\partial_\mu Y^M\partial_\nu Y^N \,=\, \eta_{AB} 
  E^A_{\ M} E^B_{\ N}\partial_\mu Y^M\partial_\nu Y^N.
  \label{ii}
\end{eqnarray}

The conversion of the curved index $M$ in the bulk into $\mu$ on the
brane can easily be done using a bulk vector tangent to the brane:
\begin{eqnarray}
  {\cal E}^A_{\ \mu} &=& E^A_{\ M} \partial_\mu Y^M.
\end{eqnarray}
This defines $d$ tangent $D$-dimensional vectors which transform
linearly under $g\in SO(1,D-1)$. What we want now 
is $e^\alpha_{\ \mu}$ which gives $d$ tangent $d$-dimensional vectors
of $SO(1,d-1)$. We already know that $\xi(\theta(x))$ converts the
$g\in G$ rotation into the $h\in H$ 
rotation; $g\xi(\theta(x))=\xi(\theta'(x))h(g,\theta(x))$. Therefore
the following quantity defined by
\begin{eqnarray}
  e^A_{\ \mu} &\equiv& \xi^{-1}(\theta(x))^A_{\ B}\, 
  {\cal E}^B_{\ \mu},
\end{eqnarray}
for each fixed $\mu$, receives only a (non-linear) $H$ rotation 
under $G$ transformation,
\begin{eqnarray}
  e'_{\,\mu} &=& \xi^{-1}(\theta'(x)){\cal E}'_{\,\mu} \,=\,
  h(g,\theta(x))\xi^{-1}(\theta(x))g^{-1}\cdot g{\cal E}_{\,\mu} \,=\,
  h(g,\theta(x))e_{\,\mu}.
\end{eqnarray}
So, $e^A_{\ \mu}$ splits into a $d$-dimensional 
vector $e^\alpha_{\ \mu}$ of $SO(1,d-1)$ and a $(D-d)$-dimensional
vector $e^a_{\ \mu}$ of $SO(D-d)$. Thus, the 
former $d$ $d$-dimensional vectors $e^\alpha_{\ \mu}$ satisfy the
property (i) which is required for our desired vielbein on the
brane. Note here that since $\xi^{-1}(\theta(x))$ itself is 
an $SO(1,D-1)$ rotation,
\begin{eqnarray}
  \eta_{AB}e^A_{\ \mu}e^B_{\ \nu} &=& \eta_{\alpha\beta} 
  e^\alpha_{\ \mu}e^\beta_{\ \nu} +\eta_{ab}e^a_{\ \mu}e^b_{\ \nu}
  \nonumber \\
  &=& \eta_{AB}{\cal E}^A_{\ \mu}{\cal E}^B_{\ \nu} \;=\; 
  \eta_{AB}E^A_{\ M}E^B_{\ N}\partial_\mu Y^M\partial_\nu Y^N.
\end{eqnarray}
Therefore, in order to satisfy the property (ii), the 
remaining $(D-d)$-dimensional vector components $e^a_{\ \mu}$ must
vanish (see, Eq.~(\ref{ii})): 
\begin{eqnarray}
  e^a_{\ \mu} &=& \xi^{-1}(\theta(x))^a_{\ B}\,{\cal E}^B_{\ \mu}
  \;=\; 0.
  \label{constraint}
\end{eqnarray}
This condition is the same as given by Sundrum~\cite{sundrum}. It
should be noted that this requirement is invariant under $G=SO(1,D-1)$
transformation since $e^a_{\ \mu}$ receives only $SO(D-d)$ rotation
under $G$. The condition (\ref{constraint}) gives a constraint on the
NG variable $\theta$. Actually, the number of equations 
in (\ref{constraint}) is $d(D-d)$ which is just the same as the number 
of $\theta$ fields. Thus the NG variables $\theta$ are completely
determined in terms 
of ${\cal E}^A_{\ \mu} =E^A_{\ M}\partial_\mu Y^M$; 
$\theta=\theta({\cal E})$. Although $\theta$ has now become dependent
variables $\theta({\cal E})$, the transformation property 
under $g\in G$ still remains the same as above. That 
is, $\theta'=\theta(g {\cal E})$ holds because the 
constraint (\ref{constraint}) is $G$-invariant. With 
this $\theta({\cal E})$, the desired induced 
vielbein $e^\alpha_{\ \mu}$ is given by
\begin{eqnarray}
  e^\alpha_{\ \mu} &=& \xi^{-1}(\theta({\cal E}))^\alpha_{\ A}\,
  {\cal E}^A_{\ \mu}.
  \label{induced}
\end{eqnarray}
It is a difficult problem to give an explicit form for the 
solution $\theta(\cal E)$. However, what we actually need is 
not $\theta(\cal E)$ but the induced vielbein $e^\alpha_{\ \mu}$. We
can find the explicit form for it as follows.

Using the explicit representation of $D$-dimensional Lorentz
generators $(J^{(AB)})^C_{\ D} =i\eta^{CE}
(\delta^A_E\delta^B_D-\delta^A_D\delta^B_E)$, the coset-space variable 
can be written as
\begin{eqnarray}
  \xi^{-1}(\theta) &=& 
  \exp\pmatrix{0 & \theta \cr -\tilde\theta & 0 \cr}
  \;=\; \bordermatrix{ & d & D-d \cr 
    \hfil d & C(\theta\tilde\theta) & S(\theta\tilde\theta)\theta \cr
    D-d & -\tilde\theta S(\theta\tilde\theta) & C(\tilde\theta\theta)
    \cr}, \\[2mm]
  &&\theta \;=\; (\theta^\alpha_{\ a}),\qquad 
  \tilde\theta^a _{\ \beta} \;\equiv\; 
  \eta^{ab}(\theta^{\rm T})_b^{\ \alpha}\eta_{\alpha\beta}, \nonumber
\end{eqnarray}
where we have introduced the functions $C(x)$ and $S(x)$, similar to
usual cosine and sine:
\begin{eqnarray}
  C(x) &\equiv& \sum_{n=0}^\infty \frac{(-1)^n}{(2n)!}x^n \;=\;
  \cos\sqrt{x}, \\
  S(x) &\equiv& \sum_{n=0}^\infty \frac{(-1)^n}{(2n+1)!}x^n \;=\;
  \frac{\sin\sqrt{x}}{\sqrt{x}}.
\end{eqnarray}
When we write the $D$-dimensional vectors ${\cal E}^A_{\ \mu}$ in the
following form splitting the parallel and transverse components to the
brane,
\begin{eqnarray}
  {\cal E}^A_{\ \mu} &=& \pmatrix{
    \Epara{\alpha}{\mu} \cr \Eperp{a}{\mu} \cr},
\end{eqnarray}
the constraint (\ref{constraint}) reads
\begin{eqnarray}
  \bigl(\tilde\theta S(\theta\tilde\theta)\bigr)^a_{\ \beta}\,
  \Epara{\beta}{\mu} &=& C(\tilde\theta\theta)^a_{\ b}\,
  \Eperp{b}{\mu},
  \label{constraint2}
\end{eqnarray}
and the induced vielbein (\ref{induced}) becomes
\begin{eqnarray}
  e^\alpha_{\ \mu} &=& C(\theta\tilde\theta)^\alpha_{\ \beta}\,
  \Epara{\beta}{\mu}
  +\bigl(S(\theta\tilde\theta)\theta\bigr)^\alpha_{\ b}\,
  \Eperp{b}{\mu}.
  \label{induced2}
\end{eqnarray}
In the above, $C(\theta\tilde\theta)^\alpha_{\ \beta}$,
$C(\tilde\theta\theta)^a_{\ b}$ and $\Epara{\alpha}{\mu}$ are all 
square matrices and invertible. Then we obtain from
Eq.~(\ref{constraint2}),
\begin{eqnarray}
  \bigl(T(\tilde\theta\theta)\tilde\theta\bigr)^a_{\ \beta} &=&
  \Eperp{a}{\mu}\,{\cal E}^{-1\,\mu}_{\parallel\ \ \ \beta}\,, 
  \label{sol1}
\end{eqnarray}
where $T(x)$ is a function analogous to tangent:
\begin{eqnarray}
  T(x) &\equiv& C^{-1}(x)S(x) \;=\; \frac{\tan\sqrt{x}}{\sqrt{x}}.
\end{eqnarray}
In deriving Eq.~(\ref{sol1}), we have used 
the ``shifting identities'' like
\begin{eqnarray}
  && \tilde\theta F(\theta\tilde\theta) \;=\;
  F(\tilde\theta\theta)\tilde\theta, \qquad
  \theta F(\tilde\theta\theta) \;=\; F(\theta\tilde\theta)\theta,
\end{eqnarray}
which generally hold for any function $F(x)$. The 
equation (\ref{sol1}) determines $\theta$ in terms of ${\cal E}$. To
obtain an expression of the induced vielbein $e^\alpha_{\ \mu}$, we
define the following quantity:
\begin{eqnarray}
  {\cal M} &\equiv& \eta ({\cal E}_\perp 
  {\cal E}^{-1}_{\parallel})^{\rm T}\eta {\cal E}_\perp 
  {\cal E}^{-1}_{\parallel} \;=\; \theta T(\tilde\theta\theta) 
  T(\tilde\theta\theta)\tilde\theta \;=\; 
  \theta\tilde\theta T^2(\theta\tilde\theta).
  \label{defcalM}
\end{eqnarray}
Using $xT^2(x)+1= C^{-2}(x)$, we further find 
\begin{eqnarray}
  C(\theta\tilde\theta) &=& (1+{\cal M})^{-1/2}.
\end{eqnarray}
Thus, the induced vielbein (\ref{induced2}) is finally rewritten as
\begin{eqnarray}
  e^\alpha_{\ \mu}&=& [C(\theta\tilde\theta)]^\alpha_{\ \gamma}
  \left(1 +(T(\theta\tilde\theta)\theta)({\cal E}_{\perp} 
    {\cal E}^{-1}_{\parallel})\right)^\gamma_{\ \beta}
  \Epara{\beta}{\mu} \nonumber \\ 
  &=& \left(1 +{\cal M}\right)^{\frac{1}{2}\,\alpha}_{\ \ \ \beta}
  \Epara{\beta}{\mu}\,.
\end{eqnarray}
This, with the definition of ${\cal M}$ in Eq.~(\ref{defcalM}), gives
the desired explicit expression for the induced vielbein. If we use
the shifting identity, we can rewrite it into a slightly more
convenient expression as
\begin{eqnarray}
  e^\alpha_{\ \mu} &=& \Epara{\alpha}{\nu}
  \bigl(1 +{\cal N}\bigr)^{1/2\> \nu}_{\ \ \ \ \;\, \mu}\,, \\[1mm]
  {\cal N} &\equiv& ({\cal E}^{\rm T}_{\parallel}\eta\, 
  {\cal E}_{\parallel})^{-1} ({\cal E}^{\rm T}_{\perp}\eta\, 
  {\cal E}_{\perp}). \nonumber
\end{eqnarray}

Finally in this Appendix, we add the expression for the induced 
spin connection $\omega_{\mu\ \,\beta}^{\ \,\alpha}(x)$ from the bulk
one $\Omega_{M\ \,B}^{\ \ \,A}(X)$. From the bulk local-Lorentz
transformation law for the covariant derivative
\begin{eqnarray}
  \partial_M+i\Omega'_{M}(X) &=& g(X)\Bigl(\partial_M+i\Omega_{M}(X)
  \Bigr)g^{-1}(X),
\end{eqnarray}
and the fact that $\xi$ is a converter of the local-Lorentz indices
from bulk to brane, it is clear that the quantity
\begin{eqnarray}
  \omega_{\mu\ \,\beta}^{\ \,\alpha}(x) &=& -i [\xi^{-1}(\partial_\mu
  +i\Omega_\mu)\xi]^\alpha_{\ \beta} \;=\; (\xi^{-1})^\alpha_{\ A}
  \Omega_{\mu\ \,B}^{\ \,A}(x)\xi^B_{\ \,\beta} -i (\xi^{-1}
  \partial_\mu\xi)^\alpha_{\ \beta}
  \label{indconn}
\end{eqnarray}
with $\Omega_{\mu\ \,B}^{\ \,A}(x)\equiv
\Omega_{M\ \,B}^{\ \ \,A}(Y(x))\partial_\mu Y^M(x)$, transforms as
\begin{eqnarray}
  \omega_\mu(x) \;\to\; \omega'_\mu(x) &=& h(x)\omega_\mu(x)h^{-1}(x) 
  -ih(x)\partial_\mu h^{-1}(x),
\end{eqnarray}
with $h$ determined by (\ref{defh}). This is just the local-Lorentz
transformation induced on the brane, and thus Eq.~(\ref{indconn})
gives the desired induced spin connection on the brane. If the bulk
connection is the usual one $\Omega_M(E)$ given in terms of the
vielbein $E_{\ \,M}^A$ as the solution 
of $\partial_{M}E_{\ \,N}^A-\Omega_{M\ \,B}^{\ \ \,A}E_{\ \,N}^{B}
-(M\leftrightarrow  N)=0$, then, the induced connection
(\ref{indconn}) is also seen to equal the usual one $\omega_\mu(e)$
given in terms of the induced vierbein $e^\alpha_{\ \,\mu}$. For
completeness, we cite the explicit expression of $\xi^{-1}$ in terms
of ${\cal E}$:
\begin{equation}
  \xi^{-1}(\theta({\cal E})) \;=\; \pmatrix{
    (1+{\cal M})^{-1/2\ \alpha}_{\qquad \ \beta} &[(1+{\cal M})^{-1/2} 
    \eta({\cal E}_\perp {\cal E}^{-1}_{\parallel})^{\rm T} 
    \eta]^\alpha_{\ \,b} \cr
    -[(1+{\cal M}')^{-1/2} {\cal E}_\perp
    {\cal E}^{-1}_{\parallel}]^a_{\ \beta} 
    & (1+{\cal M}')^{-1/2\ a}_{\qquad \ \,b} \cr}
\end{equation} 
where ${\cal M}'\equiv{\cal E}_\perp {\cal E}^{-1}_{\parallel}
\eta({\cal E}_\perp {\cal E}^{-1}_{\parallel})^{\rm T} \eta$. (Note
that $F({\cal M}'){\cal E}_\perp {\cal E}^{-1}_{\parallel}
={\cal E}_\perp {\cal E}^{-1}_{\parallel}F({\cal M})$ by shifting
identity.) $\,\xi$ is given simply by changing the signs of the
off-diagonal elements.

\section{The potential of the fifth force }

Let us first evaluate the amplitude for the diagram in Fig.~1. If the 
external matter fields are on the mass shell, only the 
part $+(1/2\tau)(\partial^\mu\phi^m\partial_\nu\phi^m)
(\bar\psi i\gamma^\nu\partial_\mu\psi)$ of the interaction 
Lagrangian (\ref{SNG}) contributes since the remaining parts vanish by
the equation of motion for $\psi$. Then the amplitude is given by
\begin{eqnarray}
  &&{\cal M} \;=\; \frac{\delta}{4\tau^2} \Bigl(\bar u(p_3)\gamma_\mu
  \frac{(p_1+p_3)_\nu}{2}u(p_1) \Bigr) \Bigl(\bar u(p_4)\gamma_\rho
  \frac{(p_1+p_3)_\sigma}{2}u(p_2) \Bigr) \times
  I^{\mu\nu\rho\sigma}(q), \nonumber \\
  &&I^{\mu\nu\rho\sigma}(q) \;=\; \int \frac{d^nk}{i(2\pi)^n}
  \frac{k^\mu k^\rho(k+q)^\nu(k+q)^\sigma +k^\nu
    k^\rho(k+q)^\mu(k+q)^\sigma}{k^2(k+q)^2},
\end{eqnarray}
where $q$ is the momentum transfer $q=p_3-p_1=p_2-p_4$ 
and $n=4-2\varepsilon$. The terms in $I^{\mu\nu\rho\sigma}(q)$, in
which any vector indices of $\mu,\,\nu,\cdots$ are carried by $q$,
cannot contribute to the amplitude because of the conservation of the
energy momentum tensor $(1/2)\bar u(p_i)\gamma_\nu(p_j+p_i)_\mu
u(p_j)$. So, the only term we have to compute is the one proportional
to $g^{(\mu\nu}g^{\rho\sigma)}\equiv g^{\mu\nu}g^{\rho\sigma}
+g^{\mu\rho}g^{\nu\sigma} +g^{\mu\sigma}g^{\rho\nu}$ which comes
solely from the $k^\mu k^\nu k^\rho k^\sigma$ term in
$I^{\mu\nu\rho\sigma}(q)$, and we find
\begin{eqnarray}
  I^{\mu\nu\rho\sigma}(q)\Big|_{g^{(\mu\nu}g^{\rho\sigma)}\, 
    {\rm term}} &=& \Gamma(\varepsilon-2) \int_0^1dx
  \left(\frac{-x(1-x)q^2}{4\pi}\right)^{2-\varepsilon}\frac{1}{2}\,
  g^{(\mu\nu}g^{\rho\sigma)} \nonumber \\
  &=& \frac{1}{64\pi^2}\,g^{(\mu\nu}g^{\rho\sigma)} \frac{(q^2)^2}{30} 
  \Bigl(C-\ln(-q^2)\Bigr),
\end{eqnarray}
where $C$ is a divergent constant
\begin{equation}
  C \;=\; \frac{1}{\bar\varepsilon} +\frac{3}{2} +\frac{47}{30},
  \qquad \Bigl(\frac{1}{\bar\varepsilon}\equiv
  \frac{1}{\varepsilon}-\gamma +\ln4\pi \Bigr).
\end{equation}
This divergent term is absorbed into a term
$g^{(\mu\nu}g^{\rho\sigma)}\Box[\bar\psi i\gamma_\mu 
({\stackrel{\leftrightarrow}{\partial}_\nu}/2)\psi]\Box [\bar\psi
i\gamma_\rho({\stackrel{\leftrightarrow}{\partial}_\sigma}/2)\psi]$
which is to appear in the higher-dimensional terms of our effective
Lagrangian. (This is the renormalization of the effective theory \`a
la Weinberg.) \ Thus the factor $\left(C-\ln(-q^2)\right)$ is replaced
by a finite one $-\ln(-q^2/\mu^2)$ with a suitable renormalization
scale $\mu$.

In the low energy limit, the amplitude is dominated by 
the $\mu=\nu=\rho=\sigma=0$ components and reduces to
\begin{equation}
  {\cal M} \;=\; \frac{mm'\delta}{4\tau^2}\frac{1}{64\pi^2}
  \left(-{3\over30}\right)(q^2)^2 \ln\Bigl(-\frac{q^2}{\mu^2}\Bigr)
  \;=\; -\frac{mm'\delta\pi^2}{160f^8}(q^2)^2
  \ln\Bigl(-\frac{q^2}{\mu^2}\Bigr),
\end{equation}
where we have used $\tau\equiv f^4/4\pi^2$ and replaced 
$\bar u(p_3)\gamma^0\bigl((p_1+p_3)^0/2\bigr)u(p_1)$ and 
$\bar u(p_4)\gamma^0\bigl((p_2+p_4)^0/2\bigr)u(p_2)$ by the matter net
masses $m$ and $m'$, respectively.

To obtain the potential, let us now evaluate the Fourier transform:
\begin{eqnarray}
  v(r) &\equiv& \int\frac{d^3q}{(2\pi)^3}\,
  e^{i{\scriptsize\mbox{\boldmath $qr$}}}
  ({\mbox{\boldmath $q$}}^2)^2 
  \ln\frac{{\mbox{\boldmath $q$}}^2}{\mu^2} \nonumber \\
  &=& \frac{1}{\pi^2}\int_0^\infty q^2dq\,
  \frac{e^{iqr}-e^{-iqr}}{iqr} q^4 \ln\frac{q^2}{\mu^2} \nonumber \\
  &=& \frac{1}{\pi^2r}\,{\rm Im}\int_0^\infty dq\, e^{iqr} q^5
  \ln(q/\mu).
\end{eqnarray}
This is divergent in the ultraviolet region $q\to\infty$. However, the
amplitude ${\cal M}\sim q^4\ln q^2$ is reliable only in the infrared
region and the true amplitude should be suppressed well in the
ultraviolet region. Supposing so, we put here a suppression 
factor $e^{-\epsilon q}$, or making a replacement $r\to r+i\epsilon$,
in the integrand and so understand that the obtained result is
reliable only for large $r$.

Then, we can rotate the integration contour $0\leq q\leq\infty$
counterclockwise into that along the imaginary 
axis, $0\leq\kappa\leq\infty$ with $q=i\kappa$ since the contribution
from the quarter circle at infinity vanishes thanks to the 
factor $e^{iqr-\epsilon q}$. Putting $\epsilon=0$, we then obtain
\begin{equation}
  v(r) \;=\; \frac{1}{\pi^2r}\,{\rm Im}\int_0^\infty d\kappa\, 
  e^{-\kappa r} \kappa^5\, i^6 \ln(i\kappa/\mu) \;=\; \frac{1}{\pi^2r} 
  \int_0^\infty d\kappa\, e^{-\kappa r} \kappa^5
  \Bigl(-\frac{\pi}{2}\Bigr) \;=\; -\frac{60}{\pi}\frac{1}{r^7}.
\end{equation}
Thus the desired potential is found to be
\begin{equation}
  V(r) \;=\; -\left(-\frac{mm'\delta\pi^2}{160f^8}\right)
  \left(-\frac{60}{\pi}\right)\frac{1}{r^7} \;=\;
  -\frac{3\pi\delta}{8f^8}\frac{mm'}{r^7},
\end{equation}
where the overall negative sign has been put since the 
amplitude ${\cal M}$ corresponds to the effective action which has
opposite sign to the potential energy.

\section{Calculation of the cross sections $\sigma_a$ and $\sigma_b$}

The amplitude for the production process (a) in Fig.~2 of the 
level $n$ KK graviton $h_{\mu\nu}^{(n)}$ is calculated using the
gravity interaction term (\ref{Sgrav}) and is given by
\begin{equation}
  {\cal M}_n \;=\; -\kappa_n\varepsilon^{*\,\mu\nu} 
  ({\mbox{\boldmath $k$}},\lambda)\,
  \bar v({\mbox{\boldmath $p$}}_2,s_2)\,\gamma_\nu
  \frac{(p_1-p_2)_\mu}{2} u({\mbox{\boldmath $p$}}_1,s_1),
\end{equation}
where $\kappa_n\equiv\kappa \exp\bigl(-\frac{1}{2}(\frac{n}{R})^2
\frac{M^2}{f^4}\bigr)$ is the $n$-th KK graviton coupling strength
suppressed by the brane fluctuation effect, and $\varepsilon_{\mu\nu}$ 
is the polarization tensor of the massive KK graviton. The
initial-spin averaged cross section in the center-of-mass frame is
given by the standard formula:
\begin{equation}
  \sigma_a \;=\; \sum_n 2\pi\delta(\sqrt s-k_0)\, \frac{\frac{1}{4}
    \sum_{s_1,s_2,\lambda}|{\cal M}_n|^2}{4s\sqrt{s-4m_e^2}},
\end{equation}
where $s=(p_1+p_2)^2$, and $\sum_n$ denotes the summation over the 
level number $n$ of the final KK graviton. Henceforth we will set the
small electron mass $m_e$ equal to zero. The spin sum for the final
massive KK graviton is done by using
\begin{equation}
  \sum_{\lambda=-2}^2 \varepsilon^{*\,\mu\nu}
  ({\mbox{\boldmath $k$}},\lambda)\, \varepsilon^{\rho\sigma}
  ({\mbox{\boldmath $k$}},\lambda) \;=\; \frac{1}{2} 
  \left(\eta^{\mu\rho}\eta^{\nu\sigma} +\eta^{\mu\sigma}\eta^{\nu\rho} 
    -\frac{2}{3}\eta^{\mu\nu}\eta^{\rho\sigma} +\cdots\right),
\end{equation}
where the ellipses denote the terms containing $k^\alpha
k^\beta/m_n^2$ ($\alpha,\beta=\mu,\nu,\rho,\sigma$), which all do not
contribute to the cross section because of the energy-momentum tensor
conservation. The third term $(2/3)\,\eta^{\mu\nu}\eta^{\rho\sigma}$
does not contribute either since the energy-momentum tensor is
traceless when $m_e=0$. Thus we obtain
\begin{equation}
  \frac{1}{4}\sum_{s_1,s_2,\lambda}|{\cal M}_n|^2 \;=\;
  \frac{\kappa_n^2}{8}s^2. 
\end{equation}
The mass of the level $n$ KK graviton is given by $m^2_n=n^2/R^2$ 
which equals $k_0^2$ in the center-of-mass frame, and so the 
summation $\sum_n$ over $n$ can be replaced by the 
integral $\int R^\delta d^\delta k_0=R^\delta\Omega_\delta\int
k_0^{\delta-1}dk_0$ with the solid 
angle $\Omega_\delta=2\pi^{\delta/2}/\Gamma(\frac{\delta}{2})$ 
in $\delta$-dimensions. Using also the 
relation $\kappa^{-2}=(2\pi R)^\delta M^{\delta+2}$, we finally get
\begin{equation}
  \sigma_a \;=\; \frac{\pi^{1-\frac{\delta}{2}}}{2^{\delta+3}
      \Gamma(\frac{\delta}{2})} \frac{s^{\delta/2}}{M^{\delta+2}}
  e^{-s\frac{M^2}{f^4}}.
\end{equation}

The amplitude for the NG boson pair production process (b) in Fig.~2
is calculated using the interaction term (\ref{SNG}) and is given by
\begin{equation}
  {\cal M}_{\phi\phi} \;=\; \frac{1}{2\tau} (-k_1^\mu k_2^\nu-k_1^\nu
  k_2^\mu)\, \bar v({\mbox{\boldmath $p$}}_2,s_2) \gamma_\nu
  \frac{(p_1-p_2)_\mu}{2} u({\mbox{\boldmath $p$}}_1,s_1). 
\end{equation}
The initial-spin averaged cross section in the center-of-mass frame 
$({\mbox{\boldmath $p$}}_1=-{\mbox{\boldmath $p$}}_2\equiv
{\mbox{\boldmath $p$}},\;{\mbox{\boldmath $k$}}_1 =
-{\mbox{\boldmath $k$}}_2\equiv {\mbox{\boldmath $k$}})$ is given by
the standard formula:
\begin{equation}
  \sigma_b \;=\; \frac{1}{2}\,\delta\int\frac{d\Omega}{64\pi^2}\,
  \frac{|{\mbox{\boldmath $k$}}|}{s\,|{\mbox{\boldmath $p$}}|}\,
  \frac{1}{4}\sum_{s_1,s_2}\left|{\cal M}_{\phi\phi}\right|^2,
\end{equation}
where the factor $1/2$ in front is put because the final two particles
are of the same kind, and the factor $\delta$ comes from the sum over
the index $m$ of the NG bosons $\phi^m$. Since we are neglecting $m_e$ 
presently, we 
have $|{\mbox{\boldmath $k$}}|=|{\mbox{\boldmath $p$}}|$, and
\begin{equation}
  \frac{1}{4}\sum_{s_1,s_2}\left|{\cal M}_{\phi\phi}\right|^2 \;=\;
  \frac{1}{128\tau^2}\,s^4\cos^2\theta(1-\cos^2\theta),
\end{equation}
with $\cos\theta\equiv {\mbox{\boldmath $p$}}\cdot 
{\mbox{\boldmath $k$}}/|{\mbox{\boldmath $p$}}|\,
|{\mbox{\boldmath $k$}}|$. Putting $\tau=f^4/4\pi^2$, we find the
cross section to be
\begin{equation}
  \sigma_b \;=\; \frac{\delta\pi^3}{1920}\frac{s^3}{f^8}.
\end{equation}

\end{document}